# Analytical Model of Modular Upper Limb Rehabilitation

M. Hasanlu[1], M. Siavashi[2]

[1]State Key Laboratory of Mechanical System and Vibration, Shanghai Jiao Tong University, Shanghai 200240, China
[2]Department of Mechanical Engineering, Babol Noshirvani University of Technology, Babol, Iran

## Abstract

Configurable robots are made up of robotic modules that can be assembled or can configure themselves into multiple robot configurations. In this research plan, a method for upper-body rehabilitation will be discussed in the form of a modular robot with different morphologies. The advantage and superiority of designing an example of a robotic module for upper body rehabilitation is the ability to reset the modular robot system. In this research, a number of modules will be designed and implemented according to the needs of one-hand rehabilitation with different degrees of freedom. The design modules' performance and efficiency will be evaluated by simulating, making samples, and testing them. This article's research includes presenting a modular upper body rehabilitation robot in the wrist, elbow, and shoulder areas, as well as providing a suitable kinematic and dynamic model of the upper body rehabilitation robot to determine human-robot interaction forces and movement. The research also involves analyzing the mathematical model of the upper body rehabilitation robot to identify advanced control strategies that rely on force control and torque control. After reviewing the articles and research of others, we concluded that no one has yet worked on the design of a prototype robotic module for upper body rehabilitation in the specified order. In our pioneering research, we intend to address this important matter.

## Introduction

The general goals of rehabilitation include the prevention of disability, the prevention of disability progression, the raising of public opinion towards the disabled and disability, the attempt to make the disabled self-sufficient and enable them to deal with problems, the adaptation of the disabled's life to society, the provision of facilities and urban services suitable for their use, the raising of awareness among the disabled about their physical and mental abilities and limitations, and the adaptation and harmonization of this situation with the environment The reconfigurable modular robot system is complex and consists of several basic module units. These basic module units can be rearranged to create different configurations to suit different environments and tasks. Generally, each module unit includes an independent driver unit, a battery unit, and a control system, and a host computer is responsible for resetting and planning the multi-module system's movement. With interfaces and actuators, it is simple to customize multi-module systems with specific functions. A new and reliable connection mechanism is designed to realize the function of connections between modules. The rehabilitation robotic module, which will be presented, is designed for the upper body. It includes a number of modules that can provide wrist movement up to three degrees of freedom. In terms of morphology, these modules are also placed in the elbow area to provide a degree of freedom for elbow movement. The third configuration is for the shoulder area, which provides shoulder movement with three degrees of freedom. In fact, a rehabilitation robot will be provided for the right or left hand area, from the shoulder to the elbow to the wrist. The reconfigurable modular robot system is complex and consists of several basic module units. These basic module units can be rearranged to create different configurations to suit different environments and tasks. Generally, each module unit includes an independent driver unit, a battery unit, and a control system, and a host computer is responsible for resetting and planning the multi-module system's movement. With interfaces and actuators, it is simple to customize multi-module systems with specific functions. We have designed a new and reliable connection mechanism to realize the functionality of connections between modules. We aim to address this important matter in our research. In an article, R.S. Calbro and colleagues described the construction and design of a robot for walking people with neurological disorders who are unable to walk properly [1]. In a descriptive article, J. Mehholz and his colleagues investigated and compared an electromechanical system to assist an injured person in walking with equipment and robots, which included endofactors and external ossification [2]. A. Koeing and colleagues, in an article, investigated and analyzed a treadmill robot that helps children with cerebral palsy walk. Their objective was to examine the dual effects of the environment on individuals with cerebral palsy. In an article, K. Sacco and colleagues presented locomotor rehabilitation mechanisms for teaching walking to patients with brain injuries. In this article, they presented their proposed model [4] with the aim of analyzing the effects of arthrosis on the patient. M.T. Karimi [5] examined the rehabilitation robots used by spinal cord injury patients in an article and discussed in detail the research done in this practical field. In an article, Z. Guo and his colleagues discussed the design of a model and mechanism for developing a skeletal robot for walking the lower limbs of the human body. Additionally, the owner of an active body weight support system, which consists of both components, can control a person who is walking [6]. A. J. Hilderley and colleagues, in an article, analyzed a robot that assists and provides walking for children with cerebral palsy by designing a functional walking program. In their article, they compared two types of RCT and RAGT approaches. These approaches were applied to walking interventions and active physical therapy for children with cerebral palsy [7]. In their descriptive article, T.G. Hornby and colleagues examined robots that help balance the body weight of people with damaged spines. Generally, based on the presentation of this article, the authors have reviewed the research and the role of these robots in patients with spinal arthritis. B. Husemann and colleagues, in an article investigating the effects of movement training to help arthritic patients after encountering an obstacle, found that this training on a walking robot was in accordance with the human movement pattern [9]. In an article, Hussain et al. designed a flexible robot for foot arthritis patients while walking on a treadmill. In this research, they used pneumatic artificial muscle to design the robot [10]. In another article, they discussed the control of the robot designed in [10]. This paper [11] employed a trajectory tracking approach, incorporating the added boundary layer sliding control theory. As well, S. Hussain et al. [12] talk about the flexible foot arthrosis robot that works with the treadmill device from [10]. They use a cooperative control approach and explain why they need an adaptive control approach to control this 6-degrees-of-freedom robot and keep it on the path they want it to follow [12]. In an article, H. Frueh and his colleagues introduced the Locomot arthritis robot and developed an educational treadmill for rehabilitation for patients whose chests were broken or traumatized; their examination methods were laboratory and experimental [13]. In a descriptive article, W.H. Chang and his colleagues discussed recent research on rehabilitation robots made for people who have suffered a stroke [14]. D.J. In an article, Reinkensneyer and colleagues investigated and optimized walking robots for humans. This descriptive article explores a range of tools, from the most basic to the newest, designed to assist patients with foot arthritis. It displays and analyzes their movement mechanisms as an example [15]. L. Lunenbuger and colleagues, in an article entitled "Walking training for people with neurological disorders," discussed and investigated various mechanisms to help people suffering from this disorder. Recent neuroscience and neurology research is more focused on finding a new way to improve the walking of people who have suffered nerve damage [16]. In another article, L. Lunenburger and colleagues investigated biological feedback in human walking rehabilitation robots. The parameters of human walking, including body weight and walking speed of human legs, are listed, and these parameters are different from the evaluation of the biofeedback resistance parameter or biological feedback [17]. In their article, J. Hilder and his colleagues discussed the kinematic tracking of the locomet robot for foot arthritis patients. This simulation yielded significant results, and it also examined the extent of motor learning in the joints and their adaptation during human walking with the robot [18]. In their article, S. Maggion et al. evaluated the robot that helps humans walk using an adaptive algorithm. In their article, Maggion et al. designed an adaptive controller that regulates the mechanical impedance of the pelvis and body weight. In this article, Hempenin conducted a laboratory study on eight healthy individuals to comprehend the human walking process and develop a training robot. to benefit patients [19]. Meyer-Heim and his colleagues, during their article research, built and paid for the clinical trial of a robot to treat children with cerebral palsy. In this research, the model designed as a virtual reality was able to increase the motivation and intensity of training for patients with disabilities in a therapeutic environment [20]. In an article, T. In underwater conditions, Miyoshi et al. investigated and developed a walking robot for foot arthritis sufferers. This robot consists of hip, knee, ankle, and foot parts. Pneumatic actuators control it, and they also form the foundation of the control approach for hip and knee angle movement [21]. D.P. Ferris et al. introduced a lower-limb robot in their article. One of the practical advantages of this robot for arthritis sufferers is that it allows them to walk, turn back, stand, and avoid obstacles in the movement path [22]. Krishnan et al., in their article, investigated the walking assistant robot. Their goal is to decrease the amount of guidance given to the robot while walking so that it is fully adapted to the patient's leg and walking performance [23]. N. In a descriptive article, Koceska and his colleagues investigated the construction of a robot that can assist and empower a patient's walking. In their article, they investigated the most important features and characteristics of rehabilitation robots. These characteristics were examined clinically as well as in terms of research and related innovations [24]. In a descriptive article, R. Riener and his colleagues investigated the virtual reality approach to rehabilitation robots for patients with arthritis. Virtual reality for rehabilitation robots has unique features. In other words, virtual reality can provide functional feedback, its multi-modal nature is beneficial for patients, and its audio and video features are among the innovations that can motivate patients [25, 26]. In a separate article, they detailed the analysis of Lokmat robots [27]. In a descriptive article, P. Sale et al. discussed the use of rehabilitation robots for people with spinal cord amputations [28]. M. In an article, A. M. Dazhir and his colleagues reviewed and described lower limb robots for patients with arthritis, as well as control approaches with pneumatic muscle actuators. Muscle agents are gentler when compared to other types of agents for foot osteoarthritis. Types of actuators can be AC, DC, pneumatic cylinders, linear actuators, series of elastic actuators and servo motors without teeth [29]. In their article, P. Wang et al. did a preliminary evaluation of rehabilitation robots for foot arthrosis sufferers. The examined robot was operated from the hip arm in a parallel position and was placed on a mobile platform [30]. P. Winchester and his colleagues, in an article, investigated the educational treadmill used to support the patient's body weight. This treadmill is equipped with a rehabilitation robot for a patient with arthritis. This robot is very useful for people with spinal cord injuries [31]. Rehabilitation robots for the upper limb usually consist of arms with different degrees of freedom, where the impact position is often graphically displayed on a computer screen, the endpoint of which is held by the patient's arm or hand. The target users are people who have suffered strokes and other neurological events, but they are not limited to the elderly, who usually suffer from immobility due to aging [32]. Likewise, the classification of upper-limb robotic systems can be seen in Table 1. In [33], the exoskeleton of the upper limb with four degrees of freedom is modeled and controlled. Because setting the PID parameters and compensating for the uncertainties of the dynamic model used in the calculated torque control can be challenging, we use time delay control to reduce or cancel the effect of uncertainties. This control is determined by a relatively simple gain selection method. The simulation results demonstrate that the proposed exoskeleton could be used as a supplement or to strengthen the user's muscle strength. Dynamic modeling and motion control of a cable-based robotic exoskeleton with pneumatic artificial muscle actuators are presented in [34]. A four-degree-of-freedom robotic exoskeleton is proposed, activated by pneumatic artificial muscle actuators, and characterized by a safe, compact, and lightweight structure, conforming to upper limb movement as much as possible. Lagrange's formula in terms of pseudo-coordinates and the principle of virtual work are used to build dynamic models that can be used for passive rehabilitation exercises. The rehabilitation path is then controlled by adaptive fuzzy sliding mode control. In [35], the modular design and control of the upper limb's exoskeleton are presented. In this article, we employ the modular design method to create the exoskeleton of the upper leg. This new design method for exoskeletons is simpler and easier to use than other techniques, and it can be developed for a larger number of articulated robots. It also provides a new admission control that functions properly in the workspace. Acceptance control is in PID format and does not require an inverse motion model or an external skeleton dynamic model. The experimental results demonstrate that both the design and the controller work well for the upper limb exoskeleton. In [36], the design and validation of the human-robot interaction system for upper extremity exoskeleton rehabilitation are presented. This paper shows how to make a motion intention detection system using a height signal sensor that will help people and exoskeleton robots work together better during upper limb rehabilitation training. A modified adaptive Kalman filter was used for the sudden change of movement mode during rehabilitation training. The results show that the recognition system designed here can correctly recognize the human-robot mutual information and estimate the human's movement intention in time. So, we can say that the designed system follows the motion predicted by the proposed method very well. This is a big step forward for controlling human-robot interactions in upper limb rehabilitation. In [37], motion and dynamic modeling of the multifunctional upper limb rehabilitation robot are presented. Understanding the precise model of a system is crucial for designing a robust and safe control system. Additionally, it can lead to a reduction in sensor costs as the model allows for the prediction of the system's output. In this context, this article focuses on identifying the motion and dynamic model of the multi-purpose rehabilitation robot, the universal tactile pantograph (UHP), and subsequently validating these models through experimental methods. The UHP is an innovative, pantograph-based robot actuated by two SEAs (Series Elastic Actuators) aimed at training the impaired upper limb after stroke. This new robot, thanks to its lockable or unlockable joints, can change its mechanical structure, enabling the stroke patient to perform various shoulder, elbow, and wrist training exercises. It focuses on the wrist pose, which is an exercise used to rehabilitate the elbow and shoulder. The kinetic model of the UHP is defined based on the vector equations, while the dynamic model is derived based on the Lagrangian formula. To demonstrate the accuracy of the models, several experimental tests were performed. The results show that the average position error between the values estimated by the model and the actual measured values remains within 3 mm (less than 2% of the maximum range of motion). Furthermore, the error between the estimated and measured force is less than 10% of the maximum force range. As a result, the developed models can be used to estimate motion, force, and control without the need for additional sensors, such as force sensors, reducing the total cost of the robot. Furthermore, the structural investigation of robotics with modular reconfiguration capability underscores the importance of the study of reference [38]. This paper reviews the current status of the development of modular reconfigurable robot systems and provides promising future research directions. So far, a wide range of modular reconfigurable robotic systems have been proposed, and these robots promise versatility, robustness, and low cost compared to other conventional robot systems. Therefore, modular reconfigurable robotic systems can perform better than traditional systems with a fixed morphology when performing tasks that require high flexibility. This paper starts by introducing the modular reconfigurable robotic classification based on its hardware architecture. It then reviews recent advances in hardware and software technologies for modular reconfigurable robotics, along with other technical issues. The traditional way of performing rehabilitation exercises has some disadvantages, including boring exercises, the impossibility of precise control over the quality of the exercises, the necessity of attendance, and limited feedback on the results of the exercises [39]. This paper introduces a configurable robot, which is composed of mathematically assembled robotic modules capable of assembling or reconfiguring themselves into various robot configurations. In this research plan, a method for upper-body rehabilitation will be discussed in the form of a modular robot with different morphologies. The advantage and superiority of designing an example of a robotic module for upper body rehabilitation is the ability to reset the modular robot system. In this research, a number of modules will be designed and implemented according to the needs of one-hand rehabilitation with different degrees of freedom. The design modules' performance and efficiency will be evaluated by simulating, making samples, and testing them. The rehabilitation robotic module, which will be presented, is designed for the upper body. It consists of multiple modules that can provide wrist movement up to three degrees of freedom. In terms of morphology, these same modules are placed in the elbow area to provide a degree of freedom for elbow movement. The third configuration is for the shoulder area, which provides shoulder movement with three degrees of freedom. In fact, a rehabilitation robot will be provided for the right or left hand area, from shoulder to elbow to wrist.

## Methodology

Kinematics is one of the first steps in industrial robot design. Kinematics allows the designer to obtain the necessary information about the position of each component in mechanical systems. This information is required for subsequent analyses of dynamics and control. In other words, kinematics is the analytical study of the robot's motion geometry relative to a fixed coordinate axis. Kinematics is examined in two modes: direct and inverse. If the parameters are obtained from the space of the joints towards the final operator, it is called direct kinematics, and if they are obtained from the side of the final operator towards the joints, it is called inverse kinematics. In complex robots, the number of answers to the kinematics problem, or simply having the answer, is very important. By using direct kinematics and the variables of the robot's joints, the position and direction of the final robot operator can be determined. The analysis of direct and inverse kinematics involves the examination of skilled mechanical arms. The variable of robot joints in hinged or rotating joints is the angle between the joints and in sliding or sliding joints, the extra length of the joint.

**Direct Kinematic Approach**

In direct kinematics, the position and direction of the robot's links and tools are calculated as a function of the jointsvariables relative to the reference axis. Coordinate devices are connected to each part of the robot mechanism, and then the relationship between these axes is expressed. The Scara robot consists of a series arm with 4 degrees of freedom, whose structure is RRPR. This arm's direct kinematic equations are as follows [40].

$$X = d_2 \cos(\theta_1 + \theta_2) + d_1 \cos \theta_1$$
$$Y = d_2 \sin(\theta_1 + \theta_2) + d_1 \sin \theta_1$$
$$Z = P$$
(1)

Equation 1 can be used to derive the speed relationship between the working space and joints.

$$\begin{bmatrix} \dot{X} \\ \dot{Y} \\ \dot{Z} \end{bmatrix} = J \begin{bmatrix} \dot{\theta}_1 \\ \dot{\theta}_2 \\ P \end{bmatrix}$$
(2)

The Jacobian matrix J is introduced. Also, based on Eq. 2, there is a solution for the problem in direct kinematic mode.

**Dynamic Model**

In the dynamics section, the goal is to find a relationship between the forces acting on the robot mechanism and their kinematic parameters. This implies that we ultimately extract the differential form of the equations of motion for further analysis; similarly to kinematics, we examine dynamics in two modes. Since the aim of this article is to explore direct kinematics and dynamics, we have omitted the description of their reverse states [40]. In direct dynamics, we determine the accelerations based on the applied forces. In reverse dynamics, forces are determined by acceleration values. Direct dynamics is mainly used for dynamic simulation, but inverse dynamics has various applications; among them, it can be used in the control line of robot movements, path design, and optimization. The following issues are of interest when solving the dynamic problem:

1. Determine the coefficients of the motion equation.

2. Identifying the inertial parameter involves estimating the inertial parameters of a robot mechanism by measuring its dynamic behavior.

3. Combined dynamics involves determining unknown forces and acceleration based on the forces in certain joints.

4. Dynamic analysis of the studied problem

A robot's motion equation can be written as follows:

$$\tau = H(q)\ddot{q} + c(q, \dot{q}, f_{ext}) \quad (3)$$

In Eq. 3, q, $\dot{q}$, $\ddot{q}$ and τ, respectively, are the vectors of position, velocity, acceleration of the joints, and torque applied to the joints. The next value represents an external force on the robot due to contact with the environment; in the present problem, it is actually the same force that must be applied to the object in order to keep it in a neutral position. The function H is referred to as the inertial matrix space, and the N*N matrix is symmetric and positive-definite. The vector C represents gravitational and coriolis terms. The robot's kinetic energy can be expressed in the following equation:

$$T = \frac{1}{2}\dot{q}^T H \dot{q} \quad (4)$$

The dynamic equation, which is mostly used for robot control purposes, can be expressed as the following relationship [1].

$$\tau = H(q)\ddot{q} + C(q,\dot{q})\dot{q} + \tau_g(q) + J(q)f_{ext} \quad (5)$$

The $\tau_g$, vector of gravitational forces $C(q,\dot{q})$, along with the vector of centrifugal and coriolis terms, $J(q)f_{ext}$ represents the overall external forces acting on the robot's joints. The previous three equations provide the equations for the body that the robot controls, as well as the equations for the robot itself. In the figure below, the terms of robot dynamics can be seen.

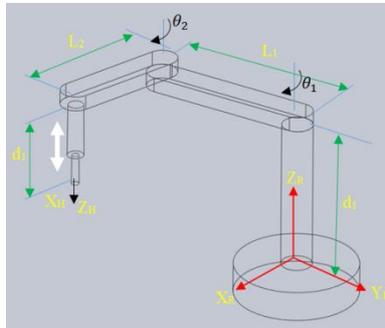

**Fig. ۴.** Schematic SCARA robot within dimension

To thoroughly investigate the problem, we consider the following assumptions for the studied system:

1. The cross-sectional area of the captured objects is uniform in such a way that the cross-sectional area of the captured area does not change due to sliding.
2. The coefficient of friction between the object and the robot is unknown, but its range is known.
3. The captured object is rigid, but the contact between the robot and the object is considered linearly elastic. This elastic property can be interpreted as a set of flexibility in either the robot or the contact surface, ultimately equating to a spring.
4. Frictional and vertical forces between the object and the robot can be measured or calculated by touch sensors embedded in the robot.
5. Each side of the object and robot has the same friction coefficient.
6. According to the above assumptions and using the Lagrange method, the robot and body equations can be obtained for the defined problem. The dynamic equation for the SCARA robot is expressed as follows:

$$T - J^T(q)f_{ext} = M(q)\ddot{q} + H(q) \quad (7)$$

The $N_r$ and $N_l$ are the vertical forces acting on the object. Note that the object's $f_{ry}$ and $f_{ly}$ forces originate from the left and right jaws. Furthermore, according to the assumption of symmetrical movement in the system, we consider only three degrees of freedom (x, y, and z) for the object. In this scenario, the problem's assumptions dictate that the body doesn't rotate. As a result, the body's motion equations are shown below.

$$\begin{bmatrix} m & 0 & 0 \\ 0 & m & 0 \\ 0 & 0 & 0 \end{bmatrix} \begin{bmatrix} \ddot{x} \\ \ddot{y} \\ \ddot{z} \end{bmatrix} = \begin{bmatrix} N_L - N_R \\ 2F_y \\ 2F_z - w \end{bmatrix} \quad (8)$$

In the above equation, $F_y$ represents the component of the frictional force applied to each of the object's left and right surfaces along the y and z axes. The parameters m and w represent the mass and weight of the object [40]. Here, we will take a practical look at the designed model to better understand the robot's 3D structure and its interactions with humans. According to the images below, the goal of presenting the 3D model is to show the robot's structure, members, and joints between them.

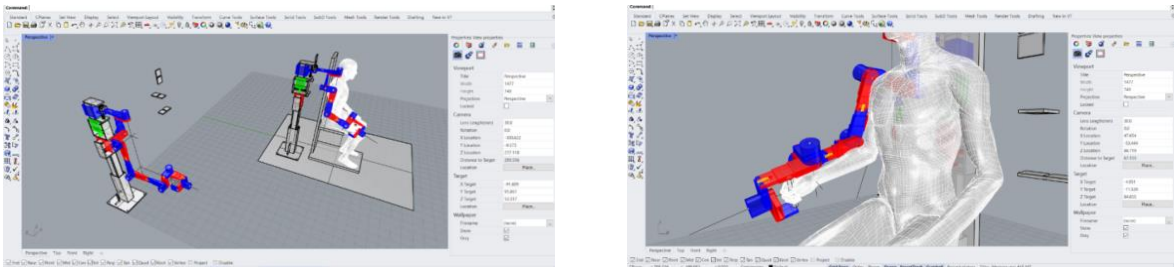

**Fig. 9.** Proposal modular robot with 3D model

## Kinematic Analysis

According to the 3D model shown in the previous images, the modular robot in question has 14 moving links and 1 fixed link (ground). This robot has 14 joints with pin (revolute) and fixed (fixed) degrees of freedom with English alphabet letters, as shown in the table below.

Table 1. Assumed joints of modular robot

| Type | Joints |
|---|---|
| Pin-motor | A, C, E, G, M, P, R |
| Fixed | B, D, F, H, N, Q, S |

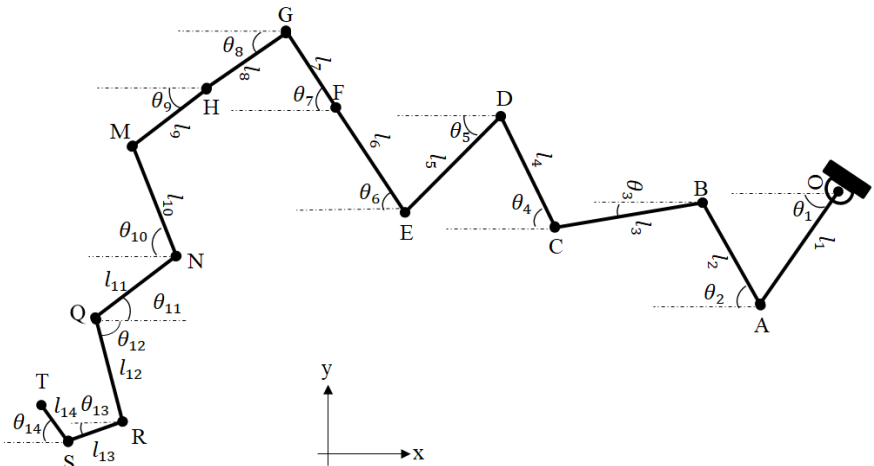

Fig. 10. Assumed mecahnism of modular robot with 15 links

To achieve kinematic analysis and degrees of freedom, we express it in Table 2. Finally, we consider the robot as a rod mechanism, making the related mathematical relationships for each joint easily understandable and presentable.

Table 2. Cartisian angular velocities and accelerations on the modular robor

| Angular acceleration | Angular velocity | axis |
|---|---|---|
| $\alpha_2$ | $\omega_2$ | X |
| $\alpha_7$ | $\omega_7$ | |
| $\alpha_4$ | $\omega_4$ | Y |
| $\alpha_5$ | $\omega_5$ | |
| $\alpha_1$ | $\omega_1$ | Z |
| $\alpha_6$ | $\omega_6$ | |

- **Joint O**

Point O is considered a fixed coordinate reference. It is obvious that this reference will not have any translational movement, so the immobile location, velocity, and acceleration are constant and zero.

Acceleration
$a_{xO} = 0$
$a_{yO} = 0$ (13)
$a_{zO} = 0$

Velocity
$v_{xO} = 0$
$v_{yO} = 0$ (12)
$v_{zO} = 0$

Coordination
$x_O = 0$
$y_O = 0$ (11)
$z_O = 0$

- **Joint T**

**Corrdination**

$$x_T = l_5 + l_1 - l_2 \cos\theta_2 + (l_6 + l_7) \sin\theta_3 + (l_8 + l_9) \cos\theta_4 + l_{10} \sin\theta_5 + l_{12} \cos\theta_6 - l_{15}$$
$$y_T = l_2 \sin\theta_1 + l_4 \cos\theta_2 + l_{11} + l_{12} \sin\theta_6 + l_{14} \cos\theta_7$$
$$z_T = l_3 + l_4 \sin\theta_2 + (l_6 + l_7) \cos\theta_3 + (l_8 + l_9) \sin\theta_4 + l_{10} \cos\theta_5 + l_{13} + l_{14} \sin\theta_7$$
(14)

**Velocity**

$$\begin{aligned}
v_{xT} &= l_2\dot\theta_1 \sin\theta_1 + (l_6+l_7)\dot\theta_3 \cos\theta_3 - (l_8+l_9)\dot\theta_4 \sin\theta_4 + l_{10}\dot\theta_5 \cos\theta_5 - l_{12}\dot\theta_6 \sin\theta_6 = l_2\omega_1 \sin\theta_1 + (l_6+l_7)\omega_3 \cos\theta_3 \\
&\quad - (l_8+l_9)\omega_4 \sin\theta_4 + l_{10}\omega_5 \cos\theta_5 - l_{12}\omega_6 \sin\theta_6 \\
v_{yT} &= l_4\dot\theta_1 \cos\theta_1 - l_4\dot\theta_2 \sin\theta_2 + l_{12}\dot\theta_6 \cos\theta_6 - l_{14}\dot\theta_7 \sin\theta_7 = l_4\omega_1 \cos\theta_1 - l_4\omega_2 \sin\theta_2 + l_{12}\omega_6 \cos\theta_6 - l_{14}\omega_7 \sin\theta_7 \\
v_{zT} &= l_4\dot\theta_2 \cos\theta_2 - (l_6+l_7)\dot\theta_3 \sin\theta_3 + (l_8+l_9)\dot\theta_4 \cos\theta_4 - l_{10}\dot\theta_5 \sin\theta_5 + l_{14}\dot\theta_7 \cos\theta_7 \\
&= l_4\omega_2 \cos\theta_2 - (l_6+l_7)\omega_3 \sin\theta_3 + (l_8+l_9)\omega_4 \cos\theta_4 - l_{10}\omega_5 \sin\theta_5 + l_{14}\omega_7 \cos\theta_7
\end{aligned} \quad (15)$$

**Acceleration**

$$\begin{aligned}
a_{xT} &= l_2\ddot\theta_1 \sin\theta_1 + l_2\dot\theta^2_1 \cos\theta_1 + (l_6+l_7)\ddot\theta_3 \cos\theta_3 - (l_6+l_7)\dot\theta^2_3 \sin\theta_3 - (l_8+l_9)\ddot\theta_4 \sin\theta_4 - (l_8+l_9)\dot\theta^2_4 \cos\theta_4 + l_{10}\ddot\theta_5 \cos\theta_5 \\
&\quad - l_{10}\dot\theta^2_5 \sin\theta_5 - l_{12}\ddot\theta_6 \sin\theta_6 - l_{12}\dot\theta^2_6 \cos\theta_6 \\
&= l_2\alpha_1 \sin\theta_1 + l_2\omega^2_1 \cos\theta_1 + (l_6+l_7)\alpha_3 \cos\theta_3 - (l_6+l_7)\omega^2_3 \sin\theta_3 - (l_8+l_9)\alpha_4 \sin\theta_4 \\
&\quad - (l_8+l_9)\omega^2_4 \cos\theta_4 + l_{10}\alpha_5 \cos\theta_5 - l_{10}\omega^2_5 \sin\theta_5 - l_{12}\alpha_6 \sin\theta_6 - l_{12}\omega^2_6 \cos\theta_6 \\
a_{yT} &= l_2\ddot\theta_1 \sin\theta_1 - l_2\dot\theta^2_1 \cos\theta_1 - l_4\ddot\theta_2 \sin\theta_2 \\
&\quad - l_4\dot\theta^2_4 \cos\theta_2 + l_{12}\ddot\theta_6 \cos\theta_6 - l_{12}\dot\theta^2_6 \sin\theta_6 - l_{14}\ddot\theta_7 \sin\theta_7 - l_{14}\dot\theta_7^{\,2} \cos\theta_7 \\
&= l_2\alpha_1 \sin\theta_1 - l_2\omega^2_1 \cos\theta_1 - l_4\alpha_2 \sin\theta_2 - l_4\omega^2_4 \cos\theta_2 + l_{12}\alpha_6 \cos\theta_6 - l_{12}\omega^2_6 \sin\theta_6 - l_{14}\alpha_7 \sin\theta_7 \\
&\quad - l_{14}\omega_7^{\,2} \cos\theta_7 \\
a_{zT} &= l_\theta\ddot\theta_1 \cos\theta_1 - l_2\dot\theta^2_1 \sin\theta_1 - (l_6+l_7)\ddot\theta_3 \sin\theta_3 - (l_6+l_7)\dot\theta^2_3 \cos\theta_3 + (l_8+l_9)\ddot\theta_4 \cos\theta_4 - (l_8+l_9)\dot\theta^2_4 \sin\theta_4 - l_{10}\ddot\theta_3 \sin\theta_3 \\
&\quad - l_{10}\dot\theta_5^{\,2} \cos\theta_5 + l_{12}\ddot\theta_6 \cos\theta_6 - l_{12}\dot\theta^2_6 \sin\theta_6 \\
&= l_\theta\alpha_1 \cos\theta_1 - l_2\omega^2_1 \sin\theta_1 - (l_6+l_7)\alpha_3 \sin\theta_3 - (l_6+l_7)\omega^2_3 \cos\theta_3 + (l_8+l_9)\alpha_4 \cos\theta_4 \\
&\quad - (l_8+l_9)\omega^2_4 \sin\theta_4 - l_{10}\ddot\theta_3 \sin\theta_3 - l_{10}\omega_5^{\,2} \cos\theta_5 + l_{12}\alpha_6 \cos\theta_6 - l_{12}\omega_6^{\,2} \sin\theta_6
\end{aligned} \quad (16)$$

## Dynamic Analysis

Basically, the purpose of examining the dynamics of any system is to examine the action and reaction of the system in the form of force and torque components. To understand the concept further, the mechanism of the modular robot is presented considering the Cartesian coordinates of the three-dimensional forces. The fundamental relations of robot dynamics are written based on Newton's second law or D'Alembert's law. Dalamber's law is used for the modular robot, and we will first review this familiar name relationship.

$$\sum M = J\ddot\theta \qquad (17)$$

If we want to simply rewrite the modular robot based on D'Alembert's law, we will have:

$$M(\theta)\ddot\theta + V(\theta,\dot\theta) + \tau_f(\theta,\dot\theta) - \lambda_u - J_1^T(\theta)\Gamma_1 - J_2^T(\theta)\Gamma_2 = \tau \qquad (18)$$

The parameters used in the previous relationship can be summarized in the following table:

**Table 3.** Dynamical parameters

| Parameter | Definition |
|---|---|
| $M(\theta)$ | Inertia matirx |
| $\theta$ | angle |
| $\dot\theta$ | Angular velocity |
| $\ddot\theta$ | Angular acceleration |
| $V(\theta,\dot\theta)$ | Central and Coriolis torques |
| $\tau_f$ | Joint friction |
| $\lambda_u$ | Error and uncertienty |
| $J_1, J_2$ | Jacobian matrice |
| $\Gamma_1, \Gamma_2$ | Inertia force |

The friction relationship is composed of two components: the friction at the joints and the skin of the members. This relationship can be expressed as follows:

$$\tau_f(\theta,\dot\theta) = \tau_r(\theta,\dot\theta) + \tau_s(\theta,\dot\theta) \qquad (19)$$

As a result, the components used in the relationship are defined in the following table.

Table 4. Frictional parameters

| Parameter | Definition |
|---|---|
| $\tau_r$ | Friction between joints |
| $\tau_s$ | Friction between links' skin |

## Result and Discussion

Displacement, speed, and acceleration are the three main components of the kinematic description of a robot. These three parameters for each joint are described in translational form, so each joint has translational displacement, translational velocity, and translational acceleration in the three Cartesian directions x, y, and z. Therefore, kinematic characteristics should be presented separately for each joint.

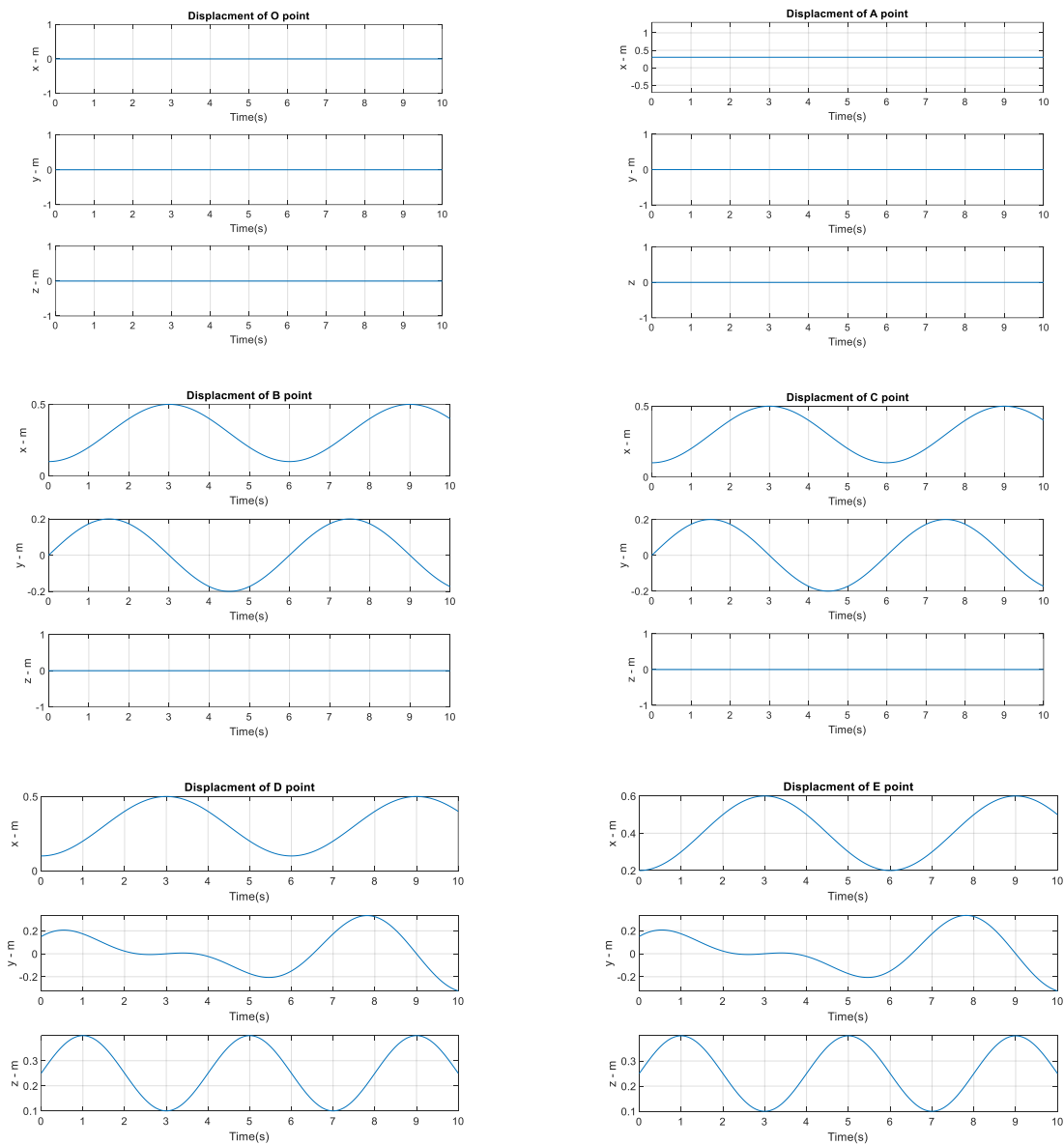

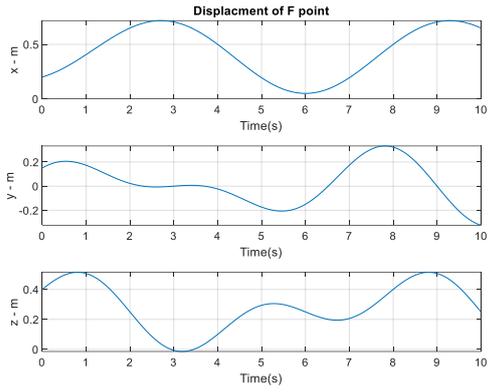
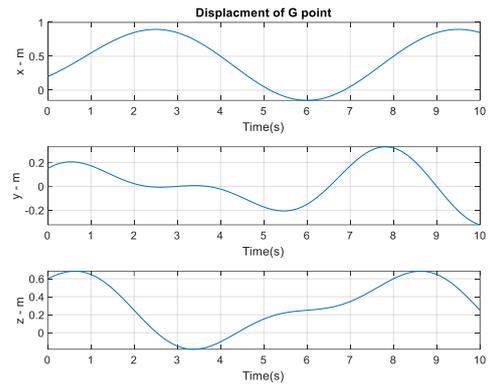
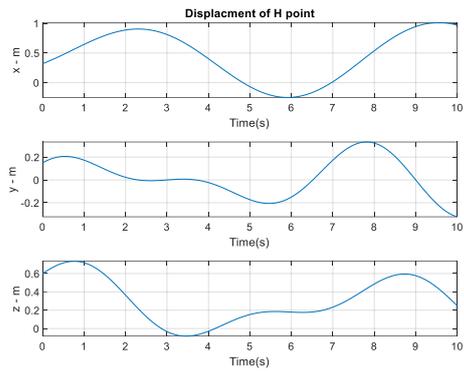
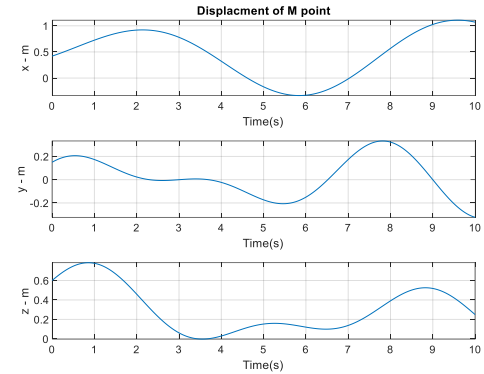
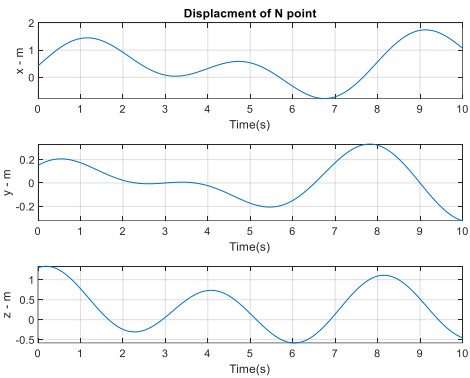
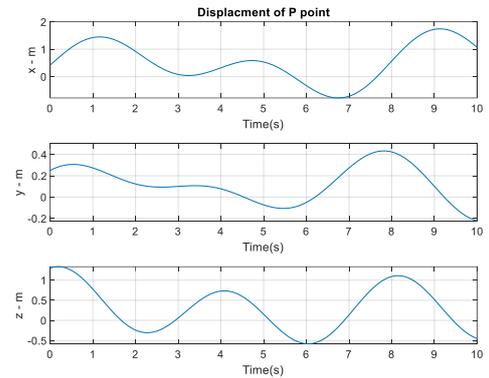
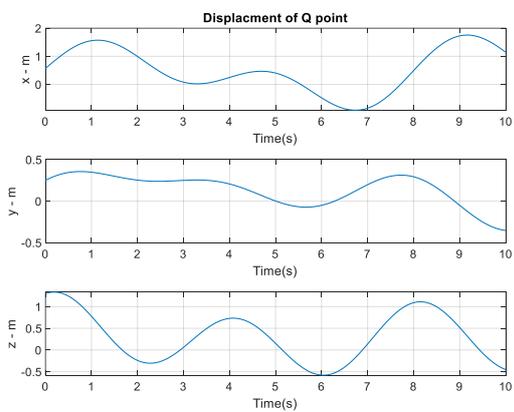
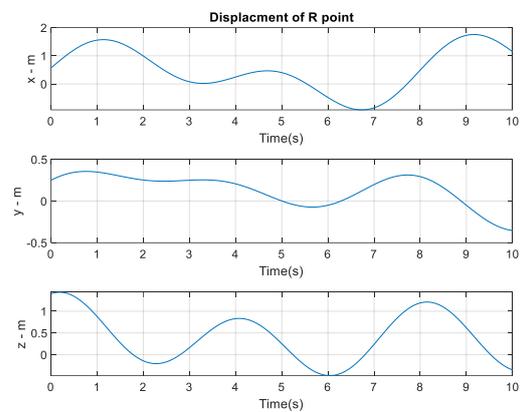

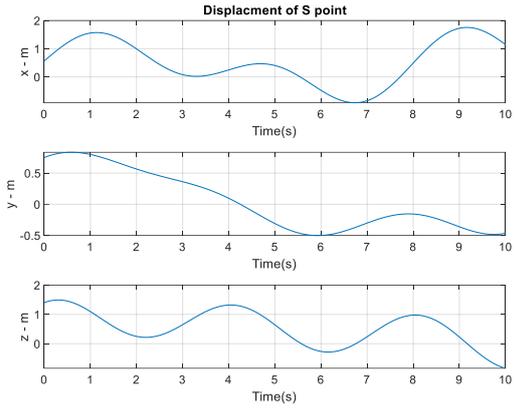
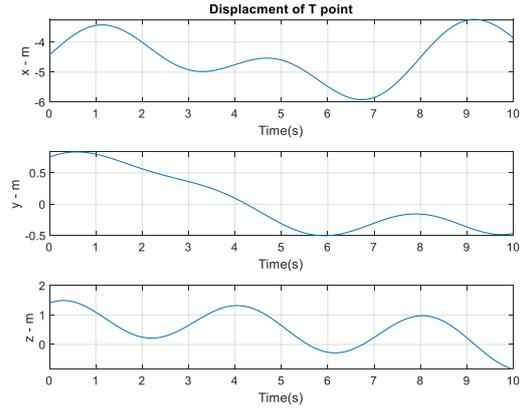

**Fig. 10.** Displacment of robot's joints

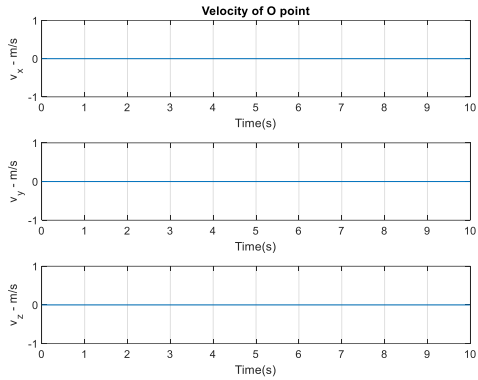
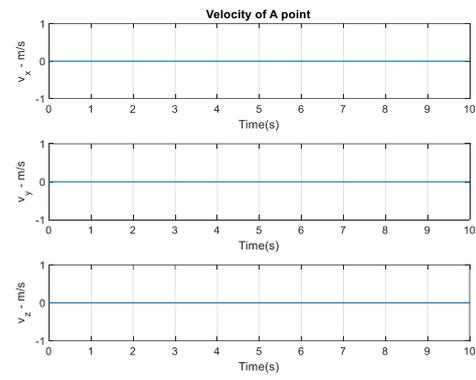
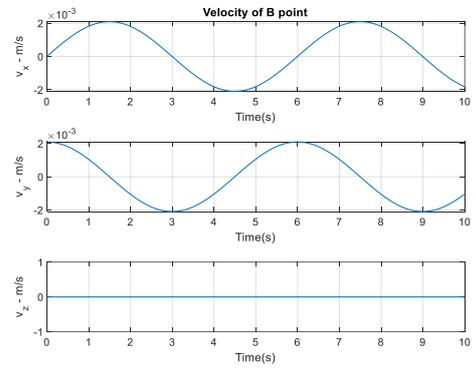
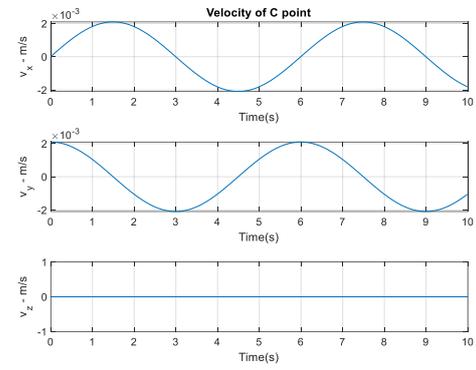
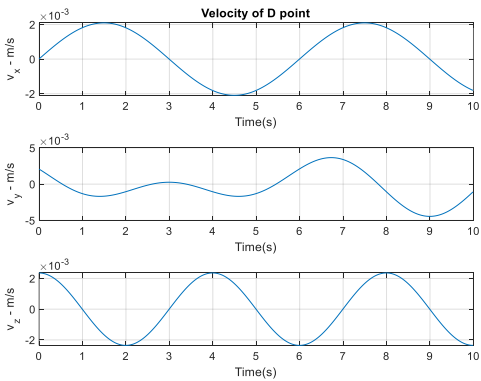
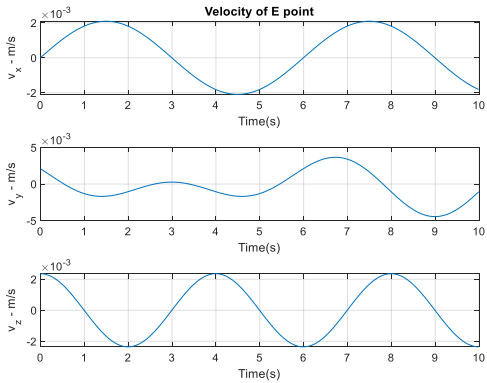

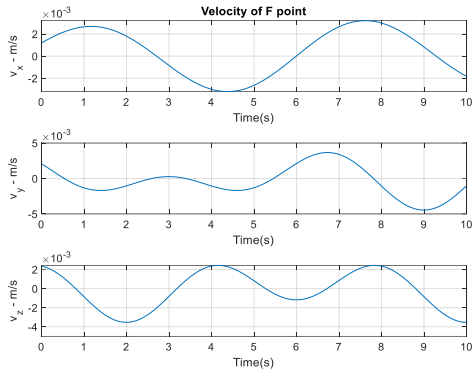
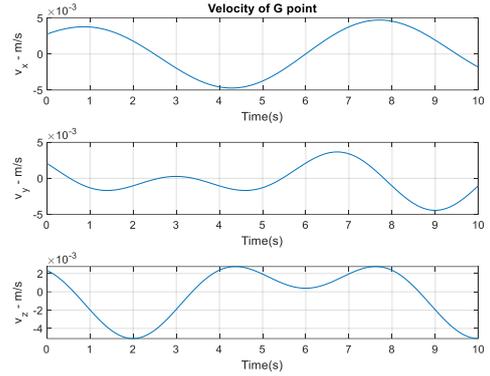
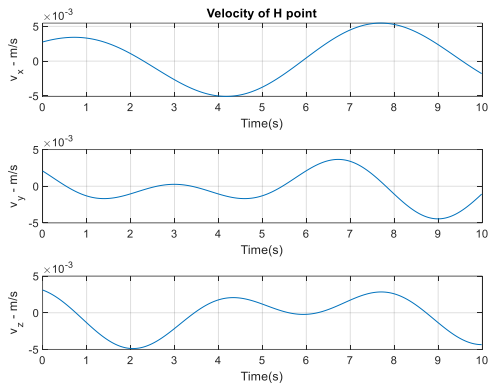
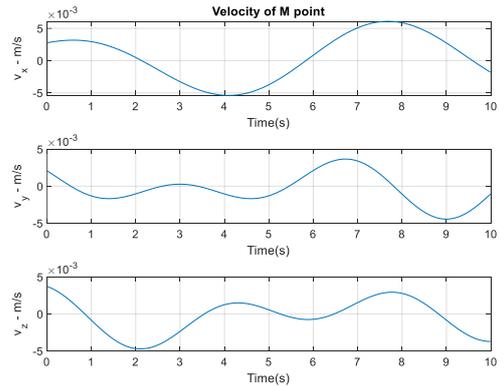
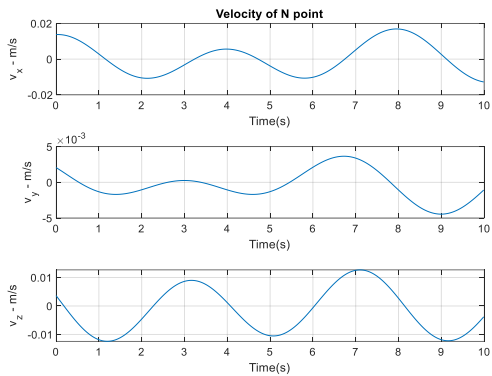
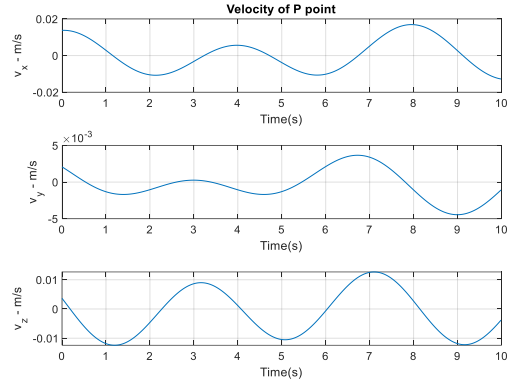
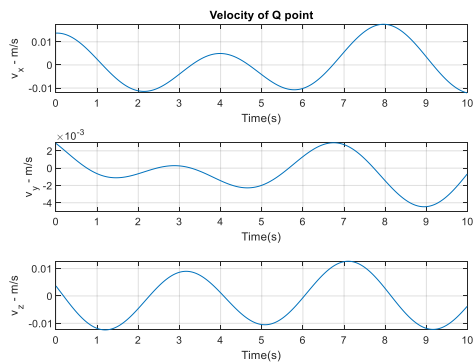
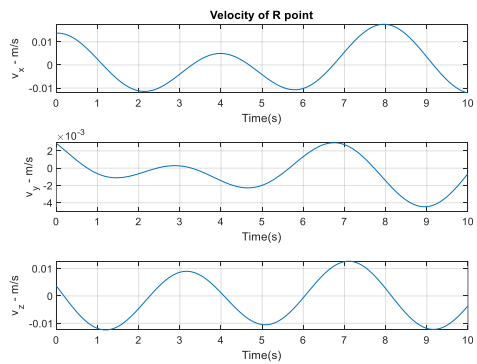

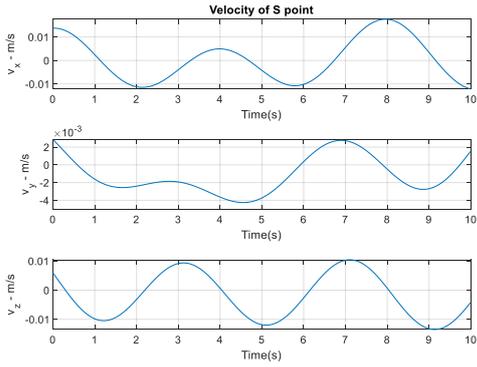
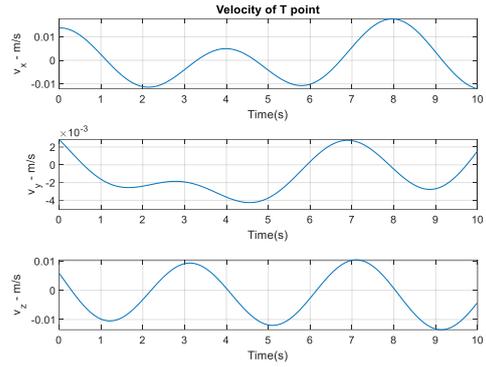

Fig. ۲۱. Angular velocity of robot's joints

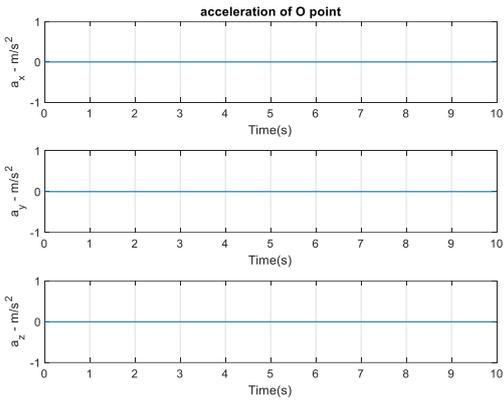
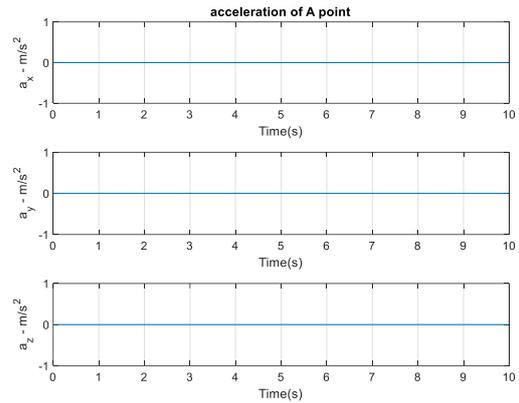

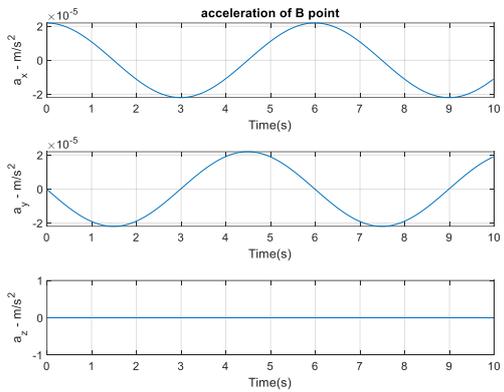
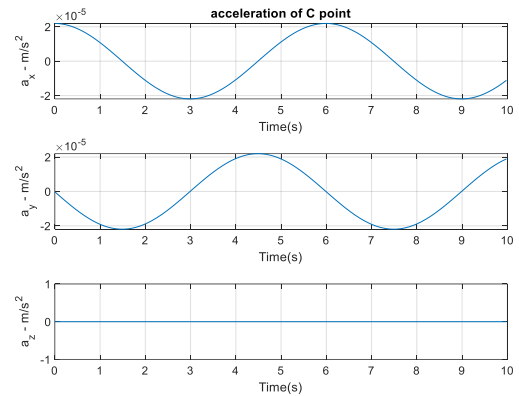

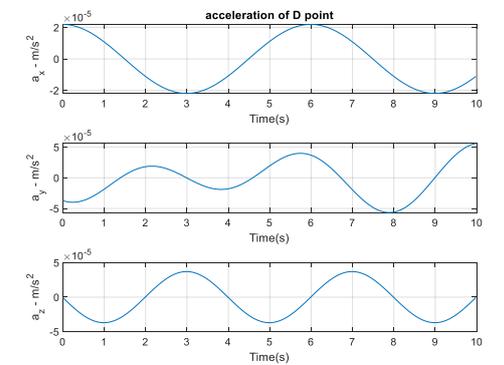
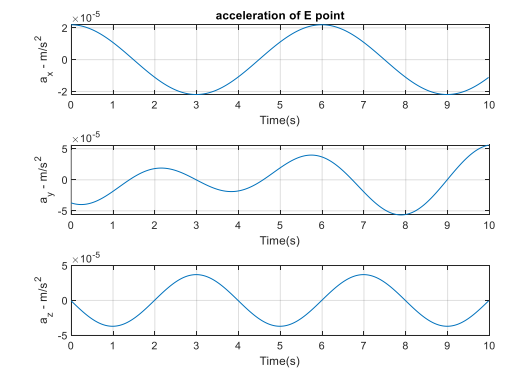

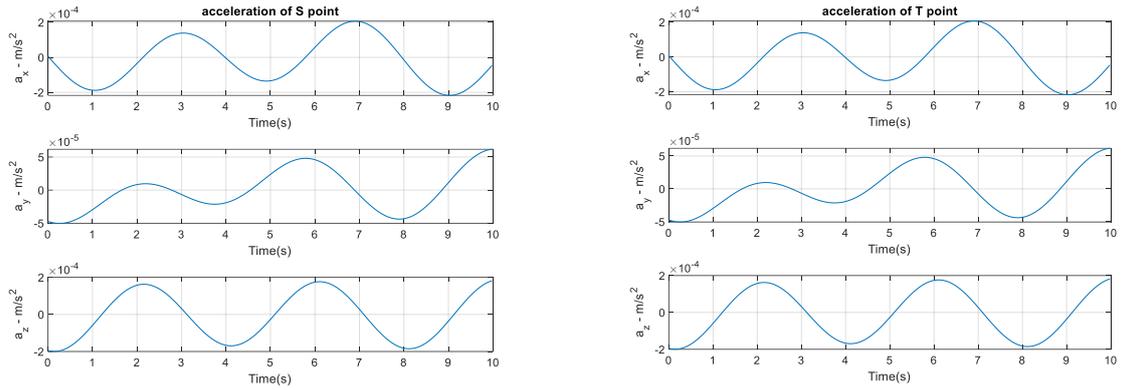

**Fig. 22.** Angular acceleration of robot's joints

In the modular robot dynamics section, it has been tried to display the moments of inertia, cryolis, and gravity of each link of the robot separately. Additionally, the following images present the equivalent torque of each robot link, based on the types of torques mentioned. As discussed in the previous section, we have replicated the mechanical and geometric properties of the modular robot for the purpose of dynamic simulation. Therefore, based on the coding done in the MATLAB software, we can divide the geometric dimensions and mass of the robot members for different people with different upper body sizes, such as children, teenagers, and adults, into small, medium, and large sizes. This division results in different results for the demonstrated robot dynamics.

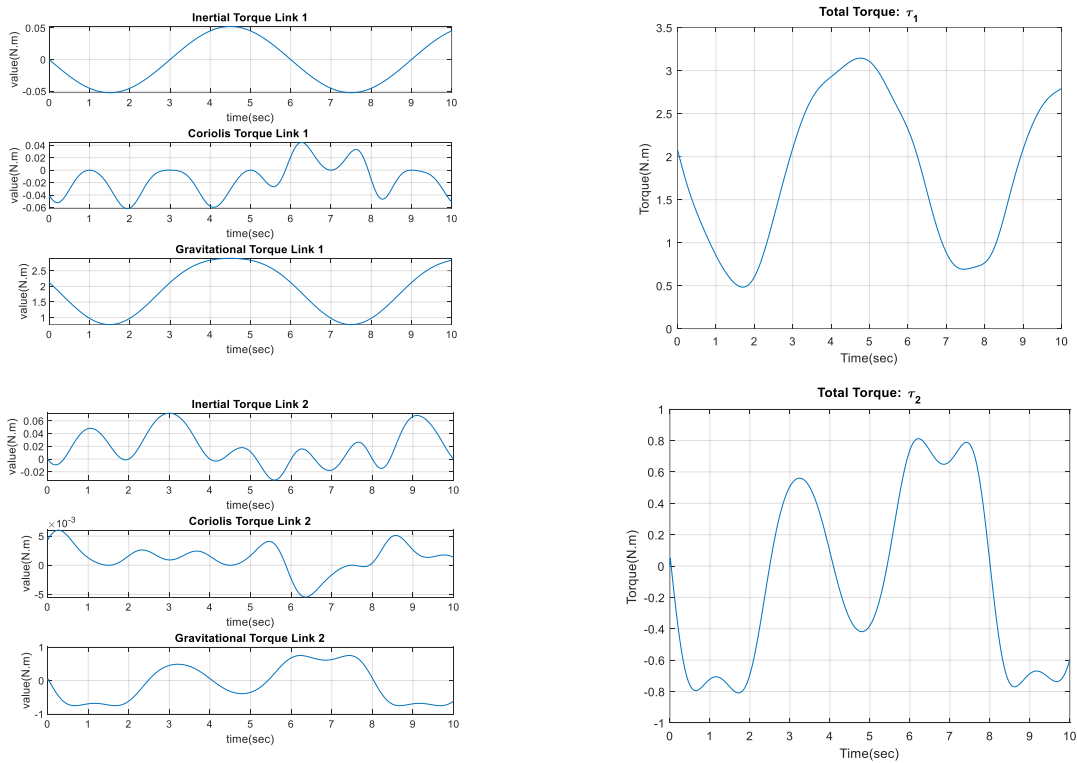

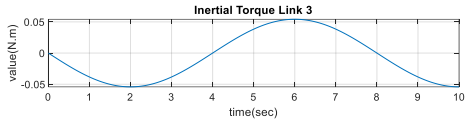
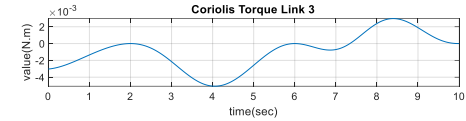
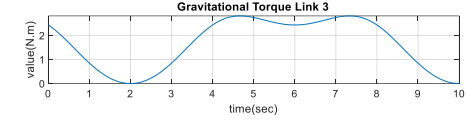
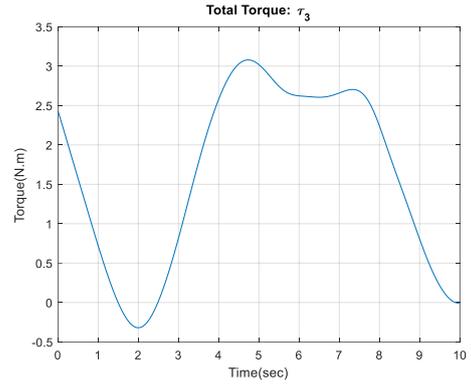
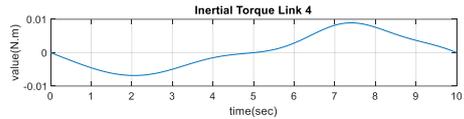
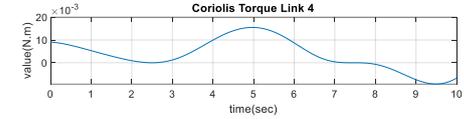
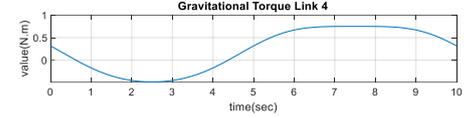
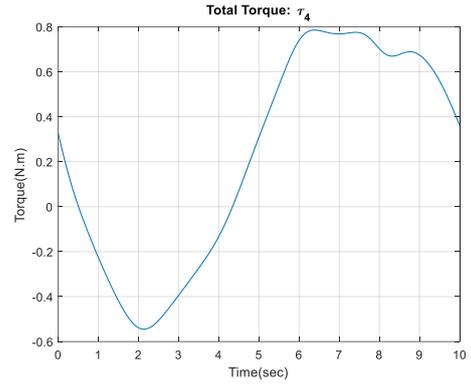
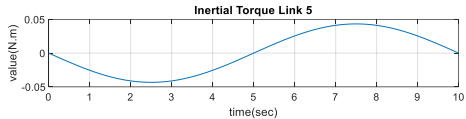
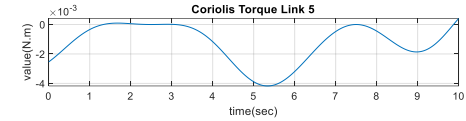
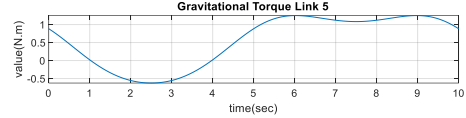
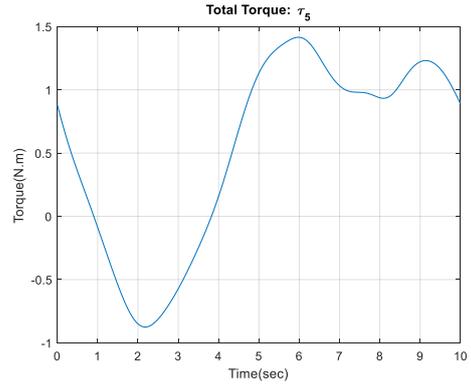
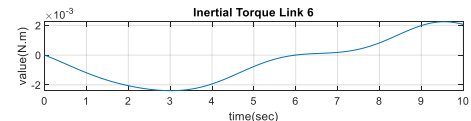
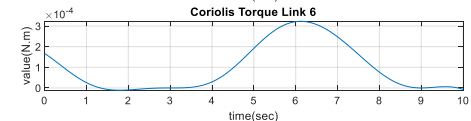
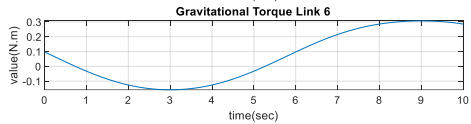
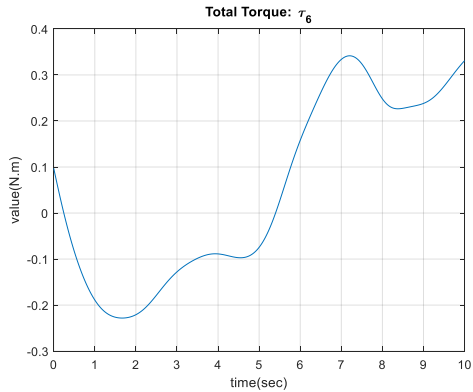

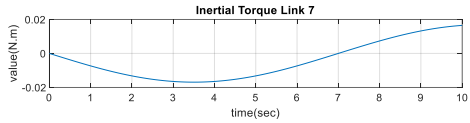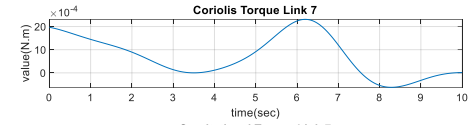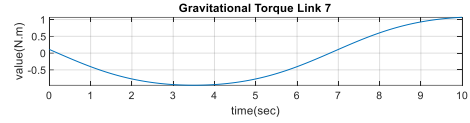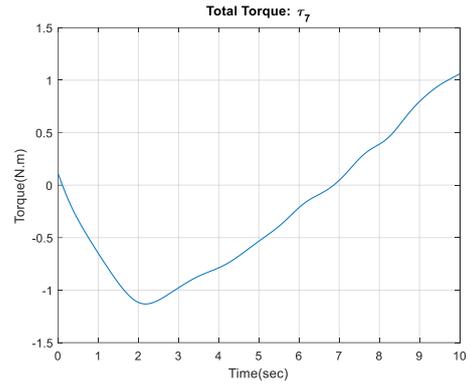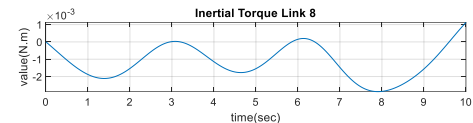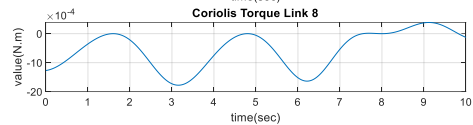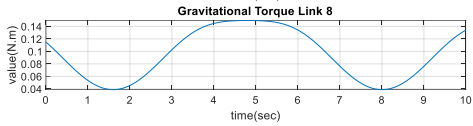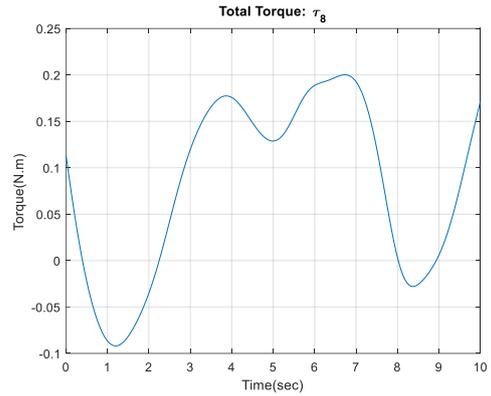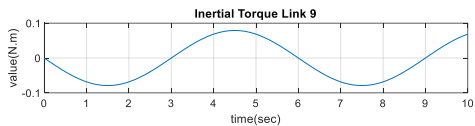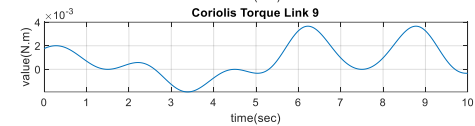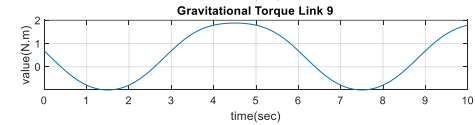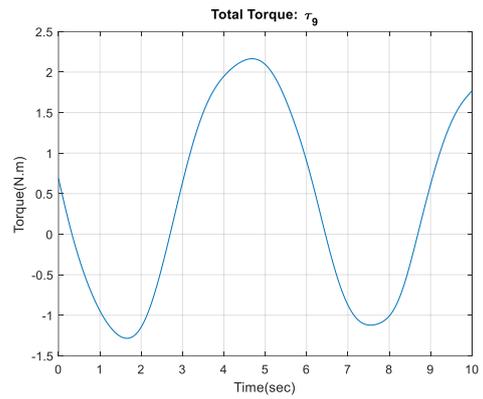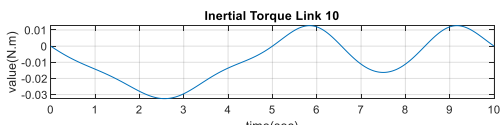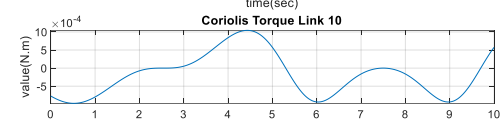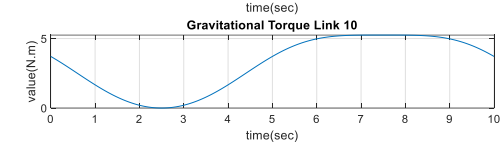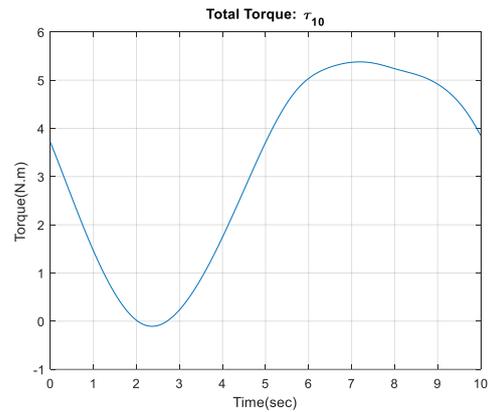

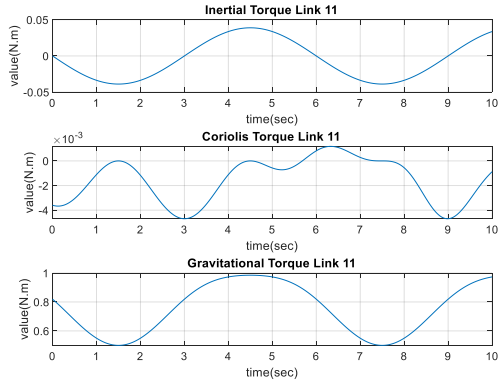
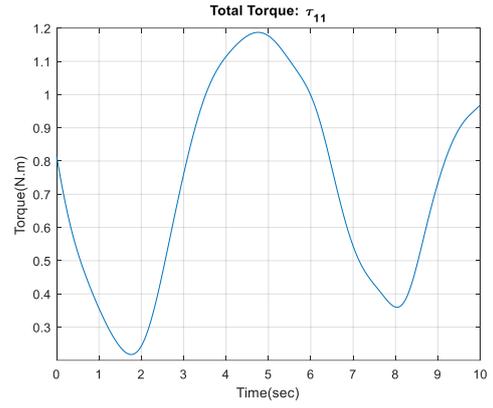
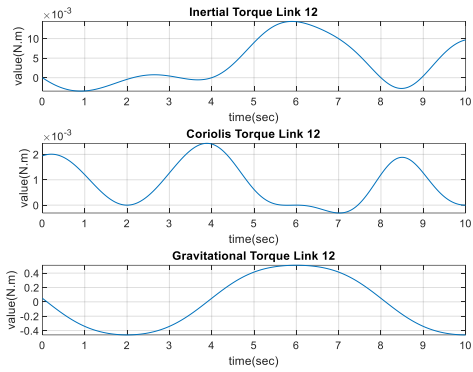
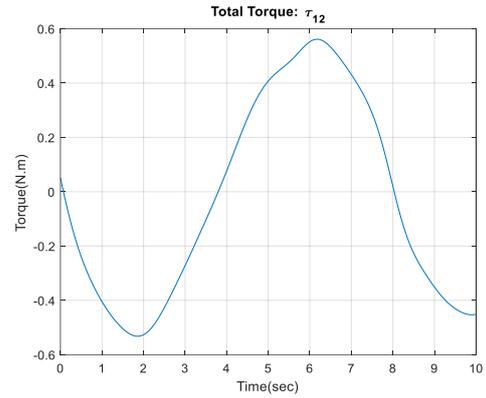
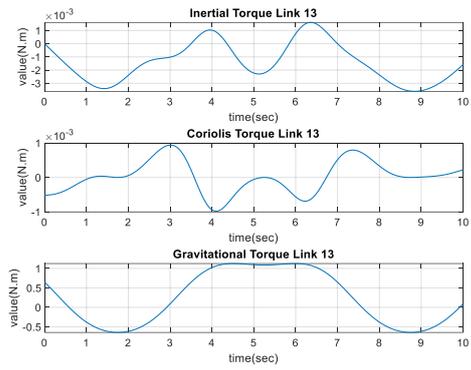
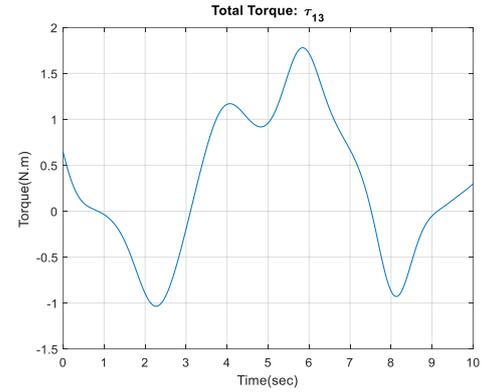
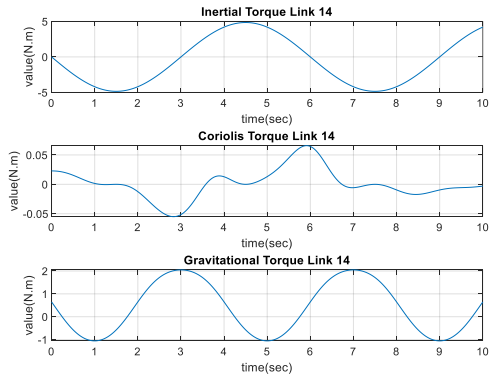
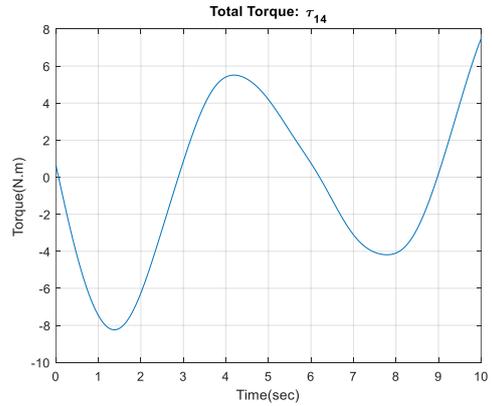

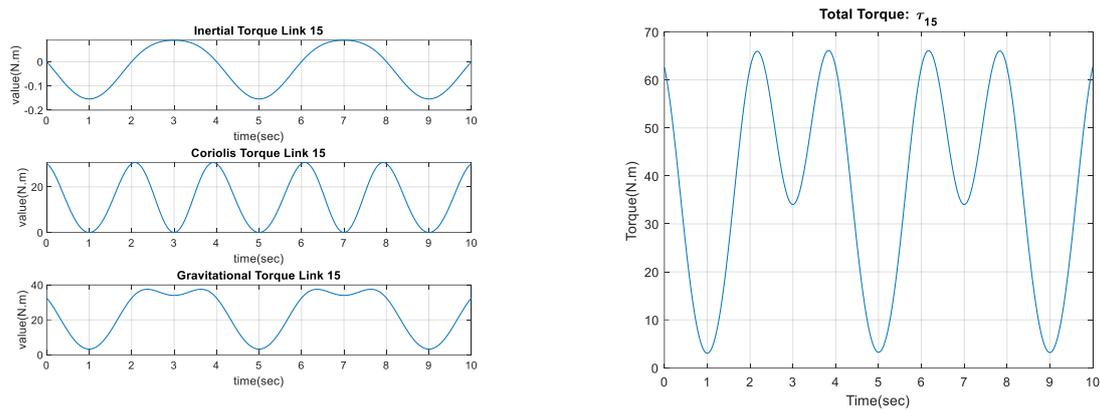

**Fig. ۲۳ .** Inertial, coriolis, gravitational, and total torques of robot's links

## Conclusion

In this article, the analysis and design of the upper body rehabilitation robotic module were researched. Medical engineering and robotics use people who have been injured in accidents and births to design and build multi-purpose, modular robots that can help disabled people. This robot's three-dimensional design was presented in the Rhino software (Figure 1) as a modular rehabilitation robot, which was presented only for the robot structure's adaptability to the human upper body. In this design, the robot was installed on the human shoulder, and the person displayed his interaction with the robot by placing his elbow, hand, and wrist on the robot. In order to present the mathematical model of this proposed robot, it was considered a 15-link mechanismTo present the mathematical model of the proposed robot, we considered it a 15-link mechanism. For each link, we formulated a three-dimensional mathematical model in the Cartesian device, representing the robot's kinematics through parameters such as displacement, speed, andNext, we attempted to code all the relations presented in the previous chapter using MATLAB computing software, and we displayed the results in full.

# Appendix

It has now initiated the kinematic modeling of the modular mechanism, focusing on points A through S. Three components of displacement, speed, and acceleration are checked for each joint. Point A, akin to a pendulum, serves as the initial moving point and the second rotational joint. We can express the mathematical relationships between its coordinate speed and translational acceleration as follows:

- **Joint A**

| Acceleration | | Velocity | | Cordination | |
|---|---|---|---|---|---|
| $a_{xA} = 0$ | (24) | $v_{xA} = 0$ | (25) | $x_A = l_1$ | (26) |
| $a_{yA} = 0$ | | $v_{yA} = 0$ | | $y_A = 0$ | |
| $a_{zA} = 0$ | | $v_{zA} = 0$ | | $z_A = 0$ | |

- **Joint B**

  Accelaration

  $a_{xB} = l_2 \alpha_1 \sin\theta_1 + l_2 \omega_1^2 \cos\theta_1$ (27)
  $a_{yB} = l_2 \alpha_1 \cos\theta_1 - l_2 \omega_1^2 \sin\theta_1$
  $a_{zB} = 0$

  Velocity

  $v_{xB} = l_2 \dot{\theta}_1 \sin\theta_1$ (28)
  $v_{yB} = l_2 \dot{\theta}_1 \cos\theta_1$
  $v_{zB} = 0$

  Cordination

  $x_B = l_1 - l_2 \cos\theta_2$ (29)
  $y_B = l_2 \sin\theta_1$
  $z_B = 0$

- **Joint C**

  Accelaration

  $a_{xC} = l_2 \alpha_1 \sin\theta_1 + l_2 \omega_1^2 \cos\theta_1$ (30)
  $a_{yC} = l_2 \alpha_1 \sin\theta_1 + l_2 \omega_1^2 \cos\theta_1$
  $a_{zC} = 0$

  Velocity

  $v_{xC} = l_1 w_1 \sin\theta_1$ (31)
  $v_{yC} = l_2 w_1 \sin\theta_1$
  $v_{zC} = 0$

  Cordination

  $x_C = l_1 - l_2 \cos\theta_2$ (32)
  $y_C = l_2 \sin\theta_1$
  $z_C = l_3$

- **Joint D**

  Accelaration

  $a_{xD} = l_2 \ddot{\theta}_1 \sin\theta_1 + l_2 \dot{\theta}_1^2 \cos\theta_1$

  $a_{yD} = l_2 \ddot{\theta}_1 \sin\theta_1 - l_2 \dot{\theta}_1 \cos\theta_1 - l_4 \ddot{\theta}_2 \sin\theta_2 - l_4 \dot{\theta}_2^2 \cos\theta_2$ (33)

  $a_{zD} = l_4 \ddot{\theta}_2 \sin\theta_2 + l_4 \dot{\theta}_2^2 \cos\theta_2$

  Velocity

  $v_{xD} = l_2 \omega_1 \sin\theta_1$

  $v_{xD} = l_2 \omega_1 \sin\theta_1$ (34)

  $v_{xD} = l_4 \omega_2 \sin\theta_1$

  Cordination

  $x_D = l_1 - l_2 \cos\theta_1$

  $y_D = l_2 \sin\theta_1 + l_4 \cos\theta_2$ (35)

  $z_D = l_3 + l_4 \sin\theta_2$

- **Joint E**

  Accelaration

  $a_{xE} = l_2 \ddot{\theta}_1 \sin\theta_1 + l_2 \dot{\theta}_1^2 \cos\theta_1$

  $a_{yE} = l_2 \ddot{\theta}_1 \sin\theta_1 - l_2 \dot{\theta}_1^2 \cos\theta_1 - l_4 \ddot{\theta}_2 \sin\theta_2 - l_4 \dot{\theta}_2^2 \cos\theta_2$ (36)

  $a_{zE} = l_2 \ddot{\theta}_2 \cos\theta_2 - l_4 \dot{\theta}_2^2 \sin\theta_2$

  Velocity

  $v_{xE} = l_2 \omega_1 \sin\theta_1$

  $v_{yE} = l_2 \omega_1 \cos\theta_1 - l_4 \omega_2 \sin\theta_2$ (37)

  $v_{zE} = l_4 \omega_2 \cos\theta_2$

  Cordination

  $x_E = l_5 + l_1 - l_2 \cos\theta_1$

  $y_E = l_2 \sin\theta_1 + l_4 \cos\theta_2$ (38)

  $z_E = l_3 + l_4 \sin\theta_2$

- **Joint F**

  Accelaration

  $a_{xF} = l_2 \ddot{\theta}_1 \sin\theta_1 + l_2 \dot{\theta}_1 \cos\theta_1 + l_6 \ddot{\theta}_3 \cos\theta_3 - l_6 \dot{\theta}_3^2 \sin\theta_3$

  $a_{yF} = l_2 \ddot{\theta}_1 \sin\theta_1 - l_2 \dot{\theta}_1^2 \cos\theta_1 - l_4 \ddot{\theta}_2 \sin\theta_2 - l_4 \dot{\theta}_2^2 \cos\theta_2$ (39)

  $a_{zF} = l_4 \ddot{\theta}_2 \cos\theta_2 - l_4 \dot{\theta}_2^2 \sin\theta_2 - l_6 \ddot{\theta}_3 \sin\theta_3 - l_6 \dot{\theta}_3^2 \cos\theta_3$

  Velocity

  $v_{xF} = l_2 \dot{\theta}_1 \sin\theta_1 + l_6 \dot{\theta}_3 \cos\theta_3$

  $v_{yF} = l_3 \dot{\theta}_1 \cos\theta_1 - l_4 \dot{\theta}_2 \sin\theta_2$ (40)

  $v_{zF} = l_4 \dot{\theta}_2 \cos\theta_2 - l_6 \dot{\theta}_3 \sin\theta_3$

  Cordination

  $x_F = l_5 + l_1 - l_2 \cos\theta_1 + l_6 \sin\theta_3$

  $y_F = l_2 \sin\theta_1 + l_4 \cos\theta_2$ (41)

  $z_F = l_3 + l_4 \sin\theta_2 + l_6 \cos\theta_3$

- **Joint G**

  Accelaration (42)       Velocity (43)       Cordination (44)

$$a_{xG} = l_2\alpha_1 \sin\theta_1 + l_2\omega_1^2 \cos\theta_1 + (l_6 + l_7)\alpha_3 \cos\theta_3 - (l_6 + l_7)\omega_3 \sin\theta_3$$

$$v_{xG} = l_2\omega_1 \sin\theta_1 + (l_6 + l_7)\omega_3 \cos\theta_3$$

$$x_G = l_5 + l_1 - l_2 \cos\theta_1 + (l_6 + l_7) \sin\theta_3$$

$$a_{yG} = l_2\alpha_1 \sin\theta_1 - l_2\omega_1^2 \cos\theta_1 - l_4\alpha_2 \sin\theta_2 - l_4\omega_2 \cos\theta_2$$

$$v_{yG} = l_2\omega_1 \cos\theta_1 - l_4\omega_2 \sin\theta_2$$

$$y_G = l_2 \sin\theta_1 + l_1 \cos\theta_2$$

$$a_{zG} = l_4\alpha_2 \cos\theta_2 - l_4\omega_2^2 \sin\theta_2 - (l_6 + l_7)\alpha_3 \sin\theta_3 - (l_6 + l_7)\omega_3^2 \cos\theta_3$$

$$v_{zG} = l_4\omega_2 \cos\theta_2 - (l_6 + l_7)\omega_3 \sin\theta_3$$

$$z_G = l_3 + l_2 \sin\theta_2 + (l_6 + l_7) \cos\theta_3$$

- **Joint H**

Cordination (45)

$$x_H = l_5 + l_1 - l_2 \cos\theta_1 + (l_6 + l_7) \sin\theta_3 + l_4 \cos\theta_4$$
$$y_H = l_2 \sin\theta_1 + l_4 \cos\theta_2$$
$$z_H = l_3 + l_4 \sin\theta_2 + (l_6 + l_7) \cos\theta_3 + l_8 \sin\theta_4$$

Velocity (46)

$$v_{xH} = l_2\omega_1 \sin\theta_1 + (l_6 + l_7)\omega_3 \cos\theta_3 - l_2\omega_4 \sin\theta_4$$
$$v_{yH} = l_2\omega_1 \cos\theta_1 - l_4\omega_2 \sin\theta_2$$
$$v_{zH} = l_4\omega_2 \cos\theta_2 - (l_6 + l_7)\omega_3 \sin\theta_3 + l_8\omega_4 \cos\theta_4$$

Accelaration (47)

$$a_{xH} = l_2\alpha_1 \sin\theta_1 + l_2\omega_1^2 \cos\theta_1 + (l_6 + l_7)\alpha_3 \cos\theta_3 - (l_6 + l_7)\alpha_3 \sin\theta_3 - l_8\omega_4 \sin\theta_4 - l_8\omega_4^2 \cos\theta_4$$
$$a_{yH} = l_2\alpha_1 \sin\theta_1 - l_2\omega_1^2 \cos\theta_1 - l_4\alpha_2 \sin\theta_2 - l_4\omega_3^2 \cos\theta_3$$
$$a_{zH} = l_4\alpha_2 \cos\theta_2 - l_4\omega_2^2 \sin\theta_2 - (l_6 + l_7)\alpha_3 \sin\theta_3 - (l_6 + l_7)\omega_3^2 \cos\theta_3 + l_8\alpha_4 \sin\theta_4 - l_8\omega_4^2 \sin\theta_4$$

- **Joint M**

Cordination (48)

$$x_M = l_5 + l_1 - l_2 \cos\theta_1 + (l_6 + l_7) \sin\theta_3 + (l_8 + l_9) \cos\theta_4$$
$$y_M = l_2 \sin\theta_1 + l_4 \cos\theta_2$$
$$z_M = l_3 + l_4 \sin\theta_2 + (l_6 + l_7) \cos\theta_3 + (l_8 + l_9) \sin\theta_4$$

Velocity (49)

$$v_{xM} = l_2 \dot\theta_1 \sin\theta_1 + (l_6 + l_7)\dot\theta_3 \cos\theta_3 - (l_8 + l_9)\dot\theta_4 \sin\theta_4$$
$$v_{yM} = l_2\dot\theta_1 \cos\theta_1 - l_4\dot\theta_2 \sin\theta_2$$
$$v_{zM} = l_4\dot\theta_2 \cos\theta_2 - (l_6 + l_7)\dot\theta_3 \sin\theta_3 + (l_8 + l_9)\dot\theta_4 \cos\theta_4$$

Accelaration (50)

$$a_{xM} = l_2\ddot\theta_1 \sin\theta_1 + l_2\dot\theta^2_1 \cos\theta_1 + (l_6 + l_7)\ddot\theta_3 \cos\theta_3 - (l_6 + l_7)\dot\theta_3^2 \sin\theta_3 - (l_8 + l_9)\ddot\theta_4 \sin\theta_4 - (l_8 + l_9)\dot\theta^2_4 \cos\theta_4$$
$$a_{yM} = l_2\ddot\theta_1 \sin\theta_1 - l_2\dot\theta^2_1 \cos\theta_1 - l_4\ddot\theta_2 \sin\theta_2 - l_4\dot\theta^2_4 \cos\theta_2$$
$$a_{zM} = l_4\ddot\theta_2 \cos\theta_2 - l_4\dot\theta^2_2 \sin\theta_2 - (l_6 + l_7)\ddot\theta_3 \sin\theta_3 - (l_6 + l_7)\dot\theta^2_3 \cos\theta_3 + (l_8 + l_9)\ddot\theta_4 \cos\theta_4 - (l_8 + l_9)\dot\theta^2_4 \sin\theta_4$$

- **Joint N**

Cordination (51)

$$x_N = l_5 + l_1 - l_2 \cos\theta_2 + (l_6 + l_7) \sin\theta_3 + (l_8 + l_9) \cos\theta_4 + l_{10} \sin\theta_5$$
$$y_N = l_2 \sin\theta_1 + l_4 \cos\theta_2$$
$$z_N = l_3 + l_4 \sin\theta_2 + (l_7 + l_8) \cos\theta_3 + (l_8 + l_9) \sin\theta_4 + l_{10} \cos\theta_5$$

Velocity (52)

$$v_{xN} = l_2\dot\theta_1 \sin\theta_1 + (l_6 + l_7)\dot\theta_3 \cos\theta_3 - (l_8 + l_9)\dot\theta_4 \sin\theta_4 + l_{10}\dot\theta_5 \cos\theta_5$$
$$v_{yN} = l_2\dot\theta_1 \cos\theta_1 - l_4\dot\theta_2 \sin\theta_2$$
$$v_{zN} = l_4\dot\theta_2 \cos\theta_2 - (l_6 + l_7)\dot\theta_3 \sin\theta_3 + (l_8 + l_9)\dot\theta_4 \cos\theta_4 - l_{10}\dot\theta_5 \sin\theta_5$$

Accelaration (53)

$$a_{xN} = l_2\ddot{\theta}_1 \sin\theta_1 + (l_6 + l_7)\dot{\theta}_3{}^2 \cos\theta_3 + (l_6 + l_7)\ddot{\theta}_3 \sin\theta_3 - (l_6 + l_7)\dot{\theta}_3{}^2 \sin\theta_3$$
$$- (l_8 + l_9)\ddot{\theta}_4 \sin\theta_4 - (l_8 + l_9)\dot{\theta}_4{}^2 \cos\theta_4 + l_{10}\ddot{\theta}_5 \cos\theta_5 - l_{10}\dot{\theta}_5{}^2 \sin\theta_5$$
$$a_{yN} = l_2\ddot{\theta}_1 \sin\theta_1 - l_2\dot{\theta}_1{}^2 \cos\theta_1 - l_4\ddot{\theta}_2 \sin\theta_2 - l_4\dot{\theta}_4{}^2 \cos\theta_2$$
$$a_{zN} = l_4\ddot{\theta}_2 \cos\theta_2 - l_4\dot{\theta}_2{}^2 \sin\theta_2 - (l_6 + l_7)\ddot{\theta}_3 \sin\theta_3 - (l_6 + l_7)\dot{\theta}_3{}^2 \cos\theta_3 + (l_6 + l_8)\ddot{\theta}_4 \cos\theta_4$$
$$- (l_8 + l_9)\dot{\theta}_4{}^2 \sin\theta_4 - l_{10}\ddot{\theta}_5 \sin\theta_5 - l_{10}\dot{\theta}_5{}^2 \sin\theta_5$$

- **Joint P**

Cordination (54)
$$x_P = l_5 + l_1 - l_2 \cos\theta_2 + (l_6 + l_7)\sin\theta_3 + (l_8 + l_9)\cos\theta_4 + l_{10}\sin\theta_5$$
$$y_P = l_2 \sin\theta_1 + l_4 \cos\theta_2 + l_{11}$$
$$z_P = l_3 + l_4 \sin\theta_2 + (l_6 + l_7)\cos\theta_3 + (l_8 + l_9)\sin\theta_4 + l_{10}\cos\theta_5$$

**Velocity** (55)
$$v_{xP} = l_2\omega_1 \sin\theta_1 + (l_6 + l_7)\omega_3 \cos\theta_3 - (l_8 + l_9)\omega_4 \sin\theta_4 + l_{10}\omega_5 \cos\theta_5$$
$$v_{yP} = l_4\omega_1 \cos\theta_1 - l_4\omega_2 \sin\theta_2$$
$$v_{zP} = l_4\omega_2 \cos\theta_2 - (l_6 + l_7)\omega_3 \sin\theta_3 + (l_8 + l_9)\omega_4 \cos\theta_4 - l_{10}\omega_3 \sin\theta_5$$

Accelaration (56)
$$a_{xP} = l_2\alpha_1 \sin\theta_1 + l_2\omega^2{}_1 \cos\theta_1 + (l_6 + l_7)\alpha_3 \cos\theta_3 - (l_6 + l_7)\omega^2{}_3 \sin\theta_3 - (l_8 + l_9)\alpha_4 \sin\theta_4$$
$$- (l_8 + l_9)\omega^2{}_4 \cos\theta_4 + l_{10}\alpha_5 \cos\theta_5 - l_{10}\omega^2{}_5 \sin\theta_5$$
$$a_{yP} = l_2\alpha_1 \sin\theta_1 - l_2\omega^2{}_1 \cos\theta_1 - l_4\alpha_2 \sin\theta_2 - l_4\omega^2{}_4 \cos\theta_2$$
$$a_{zP} = l_\theta\alpha_1 \cos\theta_1 - l_2\omega^2{}_1 \sin\theta_1 - (l_6 + l_7)\alpha_3 \sin\theta_3 - (l_6 + l_7)\omega^2{}_3 \cos\theta_3 + (l_8 + l_9)\alpha_4 \cos\theta_4$$
$$- (l_8 + l_9)\omega^2{}_4 \sin\theta_4 - l_{10}\ddot{\theta}_3 \sin\theta_3 - l_{10}\omega_5{}^2 \cos\theta_5$$

- **Joint Q**

Cordination (57)
$$x_Q = l_5 + l_1 - l_2 \cos\theta_2 + (l_6 + l_7)\sin\theta_3 + (l_8 + l_9)\cos\theta_4 + l_{10}\sin\theta_5 + l_{12}\cos\theta_6$$
$$y_Q = l_2 \sin\theta_1 + l_4 \cos\theta_2 + l_{11} + l_{12}\sin\theta_3$$
$$z_Q = l_3 + l_4 \sin\theta_2 + (l_6 + l_7)\cos\theta_3 + (l_8 + l_9)\sin\theta_4 + l_{10}\cos\theta_5$$

Velocity (58)
$$v_{xQ} = l_2\omega_1 \sin\theta_1 + (l_6 + l_7)\omega_3 \cos\theta_3 - (l_8 + l_9)\omega_4 \sin\theta_4 + l_{10}\omega_5 \cos\theta_5 - l_{12}\omega_6 \sin\theta_6$$
$$v_{yQ} = l_4\omega_1 \cos\theta_1 - l_4\omega_2 \sin\theta_2 + l_{12}\omega_6 \cos\theta_6$$
$$v_{zQ} = l_4\omega_2 \cos\theta_2 - (l_6 + l_7)\omega_3 \sin\theta_3 + (l_8 + l_9)\omega_4 \cos\theta_4 - l_{10}\omega_3 \sin\theta_5$$

Accelaration (59)
$$a_{xQ} = l_2\alpha_1 \sin\theta_1 + l_2\omega^2{}_1 \cos\theta_1 + (l_6 + l_7)\alpha_3 \cos\theta_3 - (l_6 + l_7)\omega^2{}_3 \sin\theta_3 - (l_8 + l_9)\alpha_4 \sin\theta_4$$
$$- (l_8 + l_9)\omega^2{}_4 \cos\theta_4 + l_{10}\alpha_5 \cos\theta_5 - l_{10}\omega^2{}_5 \sin\theta_5 - l_{12}\alpha_6 \sin\theta_6$$
$$- l_{12}\omega^2{}_6 \cos\theta_6$$
$$a_{yQ} = l_2\alpha_1 \sin\theta_1 - l_2\omega^2{}_1 \cos\theta_1 - l_4\alpha_2 \sin\theta_2 - l_4\omega^2{}_4 \cos\theta_2 + l_{12}\alpha_6 \cos\theta_6 - l_{12}\omega^2{}_6 \sin\theta_6$$
$$a_{zQ} = l_\theta\alpha_1 \cos\theta_1 - l_2\omega^2{}_1 \sin\theta_1 - (l_6 + l_7)\alpha_3 \sin\theta_3 - (l_6 + l_7)\omega^2{}_3 \cos\theta_3 + (l_8 + l_9)\alpha_4 \cos\theta_4$$
$$- (l_8 + l_9)\omega^2{}_4 \sin\theta_4 - l_{10}\ddot{\theta}_3 \sin\theta_3 - l_{10}\omega_5{}^2 \cos\theta_5$$

- **Joint R**

Cordination (60)
$$x_Q = l_5 + l_1 - l_2 \cos\theta_2 + (l_6 + l_7)\sin\theta_3 + (l_8 + l_9)\cos\theta_4 + l_{10}\sin\theta_5 + l_{12}\cos\theta_6$$
$$y_Q = l_2 \sin\theta_1 + l_4 \cos\theta_2 + l_{11} + l_{12}\sin\theta_6$$
$$z_Q = l_3 + l_4 \sin\theta_2 + (l_6 + l_7)\cos\theta_3 + (l_8 + l_9)\sin\theta_4 + l_{10}\cos\theta_5$$

Velocity (61)
$$v_{xR} = l_2\omega_1 \sin\theta_1 + (l_6 + l_7)\omega_3 \cos\theta_3 - (l_8 + l_9)\omega_4 \sin\theta_4 + l_{10}\omega_5 \cos\theta_5 - l_{12}\omega_6 \sin\theta_6$$
$$v_{yR} = l_4\omega_1 \cos\theta_1 - l_4\omega_2 \sin\theta_2 + l_{12}\omega_6 \cos\theta_6$$
$$v_{zR} = l_4\omega_2 \cos\theta_2 - (l_6 + l_7)\omega_3 \sin\theta_3 + (l_8 + l_9)\omega_4 \cos\theta_4 - l_{10}\omega_3 \sin\theta_5$$

Accelaration (62)
$$a_{xR} = l_2\alpha_1 \sin\theta_1 + l_2\omega^2{}_1 \cos\theta_1 + (l_6 + l_7)\alpha_3 \cos\theta_3 - (l_6 + l_7)\omega^2{}_3 \sin\theta_3 - (l_8 + l_9)\alpha_4 \sin\theta_4$$
$$- (l_8 + l_9)\omega^2{}_4 \cos\theta_4 + l_{10}\alpha_5 \cos\theta_5 - l_{10}\omega^2{}_5 \sin\theta_5 - l_{12}\alpha_6 \sin\theta_6$$
$$- l_{12}\omega^2{}_6 \cos\theta_6$$
$$a_{yR} = l_2\alpha_1 \sin\theta_1 - l_2\omega^2{}_1 \cos\theta_1 - l_4\alpha_2 \sin\theta_2 - l_4\omega^2{}_4 \cos\theta_2 + l_{12}\alpha_6 \cos\theta_6 - l_{12}\omega^2{}_6 \sin\theta_6$$
$$a_{zR} = l_\theta\alpha_1 \cos\theta_1 - l_2\omega^2{}_1 \sin\theta_1 - (l_6 + l_7)\alpha_3 \sin\theta_3 - (l_6 + l_7)\omega^2{}_3 \cos\theta_3 + (l_8 + l_9)\alpha_4 \cos\theta_4$$
$$- (l_8 + l_9)\omega^2{}_4 \sin\theta_4 - l_{10}\ddot{\theta}_3 \sin\theta_3 - l_{10}\omega_5{}^2 \cos\theta_5$$

- **Joint S**

Cordination (63)
$$x_S = l_5 + l_1 - l_2 \cos\theta_2 + (l_6 + l_7)\sin\theta_3 + (l_8 + l_9)\cos\theta_4 + l_{10}\sin\theta_5 + l_{12}\cos\theta_6$$
$$y_S = l_2 \sin\theta_1 + l_4 \cos\theta_2 + l_{11} + l_{12}\sin\theta_6 + l_{14}\cos\theta_7$$

$$z_S = l_3 + l_4 \sin \theta_2 + (l_6 + l_7) \cos \theta_3 + (l_8 + l_9) \sin \theta_4 + l_{10} \cos \theta_5 + l_{13} + l_{14} \sin \theta_7$$

(64)

Velocity

$$v_{xS} = l_2 \dot{\theta}_1 \sin \theta_1 + (l_6 + l_7)\dot{\theta}_3 \cos \theta_3 - (l_8 + l_9)\dot{\theta}_4 \sin \theta_4 + l_{10}\dot{\theta}_5 \cos = l_2 \omega_1 \sin \theta_1 \\ + (l_6 + l_7)\omega_3 \cos \theta_3 - (l_8 + l_9)\omega_4 \sin \theta_4 + l_{10}\omega_5 \cos \theta_5 - l_{12}\omega_6 \sin \theta_6$$

$$v_{yS} = l_4 \omega_1 \cos \theta_1 - l_4 \omega_2 \sin \theta_2 + l_{12}\omega_6 \cos \theta_6 - l_{14}\omega_7 \sin \theta_7$$

$$v_{zS} = l_4 \omega_2 \cos \theta_2 - (l_6 + l_7)\omega_3 \sin \theta_3 + (l_8 + l_9)\omega_4 \cos \theta_4 - l_{10}\omega_5 \sin \theta_5 + l_{14}\omega_7 \cos \theta_7$$

(65)

Accelaration

$$a_{xS} = l_2 \alpha_1 \sin \theta_1 + l_2 \omega^2{}_1 \cos \theta_1 + (l_6 + l_7)\alpha_3 \cos \theta_3 - (l_6 + l_7)\omega^2{}_3 \sin \theta_3 - (l_8 + l_9)\alpha_4 \sin \theta_4 \\ - (l_8 + l_9)\omega^2{}_4 \cos \theta_4 + l_{10}\alpha_5 \cos \theta_5 - l_{10}\omega^2{}_5 \sin \theta_5 - l_{12}\alpha_6 \sin \theta_6 \\ - l_{12}\omega^2{}_6 \cos \theta_6$$

$$a_{yS} = l_2 \alpha_1 \sin \theta_1 - l_2 \omega^2{}_1 \cos \theta_1 - l_4 \alpha_2 \sin \theta_2 - l_4 \omega^2{}_4 \cos \theta_2 + l_{12}\alpha_6 \cos \theta_6 - l_{12}\omega^2{}_6 \sin \theta_6 \\ - l_{14}\alpha_7 \sin \theta_7 - l_{14}\omega_7{}^2 \cos \theta_7$$

$$a_{zS} = l_\theta \alpha_1 \cos \theta_1 - l_2 \omega^2{}_1 \sin \theta_1 - (l_6 + l_7)\alpha_3 \sin \theta_3 - (l_6 + l_7)\omega^2{}_3 \cos \theta_3 + (l_8 + l_9)\alpha_4 \cos \theta_4 \\ - (l_8 + l_9)\omega^2{}_4 \sin \theta_4 - l_{10}\ddot{\theta}_3 \sin \theta_3 \\ - l_{10}\omega_5{}^2 \cos \theta_5 + l_{12}\alpha_6 \cos \theta_6 - l_{12}\omega_6{}^2 \sin \theta_6$$